\documentclass[english]{article}
\usepackage[T1]{fontenc}
\usepackage{url}
\usepackage[utf8]{inputenc}
\usepackage[english]{babel}
\usepackage{geometry}
\usepackage{float}
\usepackage{amsmath}
\usepackage{graphicx}
\usepackage{xcolor}
\usepackage{lipsum}
\usepackage{natbib}
\makeatletter

\floatstyle{ruled}
\newfloat{algorithm}{tbp}{loa}
\providecommand{\algorithmname}{Algorithm}
\floatname{algorithm}{\protect\algorithmname}
\usepackage{lineno}

\makeatother

\usepackage{babel}
\begin{document}

\title{Spectral Properties of Solar Wind Charge Exchange from Earth’s Magnetospheric Boundaries: MHD and Test Particle Simulations}

\author{A. Ramachandran$^{1}$\thanks{corresponding author: ardra.kozhikottuparambil@warwick.ac.uk}, R. T. Desai$^{1,2}$, H. L. Norman$^{1}$, Y. Tong$^{1}$, D. Bewicke$^{1}$,\\ S. Nitti$^{3}$, J. A. Carter$^{4}$, J. P. Eastwood$^{2}$, J. P. Chittenden$^{2}$}

\date{}
\maketitle
\vspace{-2em}

\begin{center}
$^1$Centre for Fusion, Space \& Astrophysics, University of Warwick, Coventry, UK\\[0.3em]
$^2$Blackett Laboratory, Imperial College London, London, UK\\[0.3em]
$^3$School of Physics and Astronomy, University of Leicester, Leicester, UK\\[0.3em]
$^4$Space Park Leicester, Corporation Road, Space City, Leicester, UK
\end{center}

\vspace{-1em}
\begin{abstract}

The Solar Wind Charge Exchange (SWCX) mechanism, where X-rays can be generated through interactions between heavy, highly charged solar wind ions and neutral atoms, offers an opportunity to observe the large-scale transfers of energy and momentum through the Sun–Earth system. In this study, we simulate SWCX signatures from Earth’s outer boundaries using Gorgon global magnetohydrodynamic (MHD) simulations combined with kinetic particle-based simulations which we call the X‑ray Imaging of the Magnetosphere (XIM) framework. 
The XIM-predicted emissivities agree with the predictions of the MHD model, in producing enhanced emissions from the magnetosheath and cusp regions, but several differences are apparent. 
Compared with MHD, XIM shows enhanced structuring in the magnetosheath and notable enhanced emissions along the magnetopause associated with betatron acceleration. The results also evidence species-specific spatial variations, which are shown to vary compared to anticipated relative abundances in the solar wind and ion-specific properties. These simulations provide new insights into the underlying dynamics of SWCX at the Earth's outer boundaries, which can be used to understand what might be observed using X-ray observatories in space, such as the recently launched SMILE mission. 
\end{abstract}

\section{Introduction}

The Earth's magnetosphere is a cavity in the solar wind, carved out by the Earth's dipolar magnetic field. The large-scale structure and temporal variations of magnetospheric boundaries, such as the magnetopause, magnetosheath, cusps, and bow shock, have so far been constrained primarily through in situ observations from either individual or multi-spacecraft missions \citep[e.g.][]{Heppner1963,Escoubet2001}. While repeated observations can statistically constrain the large-scale nature of the system \citep[e.g.][]{Shue1998,Lin2010}, global imaging via the solar wind charge exchange (SWCX) process \citep{cravens1997swcx, cravens2001swcx}, offers an alternative direct method to instantaneously resolve the large-scale magnetospheric structure and dynamics in response to solar wind driving \citep{sibeck2018imaging, sibeck2023quantifying}. 

SWCX occurs between highly charged solar wind ions and neutral atoms in Earth's hydrogen exosphere and can be expressed by the reaction:
\begin{equation}\label{eq_1}
M+X^{q+}\rightarrow M^++X^{(q-1)+*}\rightarrow M^++X^{(q-1)+}+\gamma_j,
\end{equation}
where a solar wind ion, \(X^{q+}\), captures an electron from a neutral target, \(M\), producing an excited ion, \(X^{(q-1)+*}\), which subsequently de-excites by emitting an X-ray photon, \(\gamma_j\), that can be measured \citep{koutroumpa2023solarwindioncharge}. Soft X-rays in the 0.1--2.0 keV energy range are produced primarily by highly ionized solar wind ions, such as \(O^{6+}\), \(O^{7+}\), and \(C^{5+}\), interacting with Earth's neutral hydrogen exosphere. These ions, originating from the solar atmosphere, primarily populate the solar wind, magnetosheath, and cusp regions and generally do not penetrate deep into the magnetosphere in abundance \citep{von2000composition, schwadron2000implications, rakhmanova2016correlation}. As a result, SWCX soft X-ray emission is predominantly produced along the magnetospheric boundaries and makes SWCX emission a useful tracer of phenomena occurring in these regions \citep{wang2022methods}.

SWCX emission has been observed throughout the Solar System, including at the Moon \citep{collier2014lunar}, Venus \citep{dennerl2008venus, dennerl2002venus}, Mars \citep{dennerl2002discovery, gunell2004x}, Jupiter \citep{dunn2022jupiter}, and comets \citep[e.g.,][]{cravens2002, krasnopolsky2004x}. Geocoronal soft X-ray emission was first identified using observations from the ROSAT (ROentgen SATellite) mission and has since been extensively studied with astronomical X-ray observatories, including XMM-Newton \citep{TRUMPER1982241, cravens2001swcx, jansen2001xmm, snowden2004xmm}. Using XMM-Newton observations, \citet{jenny2009XMM} and \citet{carter2011identifying} observed soft X-ray emission from near-Earth regions and concluded that the dayside magnetosheath is a major source of geocoronal SWCX emission. Later, \citet{whittaker2016modeling} used XMM-Newton observations to validate MHD simulations of magnetosheath SWCX emission, while \citet{nittin2024swtype} used these observations to investigate different solar wind types during periods of geocoronal SWCX. These near-Earth SWCX observations provided additional motivation for the development of dedicated missions such as LEXI (Lunar Environment Heliospheric X-ray Imager \citep{walsh2024lunar}) and SMILE (Solar Wind Magnetosphere Ionosphere Link Explorer; \citep{wang2018progress, philippe2025solar}).

SMILE aims to image Earth's magnetosphere using soft X-rays. It is a collaborative mission between the European Space Agency (ESA) and the Chinese Academy of Sciences (CAS), launched in May 2026. The Soft X-ray Imager (SXI), equipped with CCD detectors, is a wide-field lobster-eye telescope with a field of view (FOV) of \(16^\circ \times 27^\circ\). It will conduct soft X-ray observations of the subsolar magnetopause and cusp regions \citep{sembay2024soft, sembay2025soft}. In contrast, previous astronomical X-ray missions, such as XMM-Newton, had relatively narrow fields of view and therefore captured only a limited portion of the near-Earth environment \citep{connor2021soft}. As the spacecraft orbits Earth, the SXI field of view continuously changes, so as to produce two-dimensional images of the magnetosphere. However, reconstructing the three-dimensional magnetopause from these images remains a significant challenge. Several methods, including the Tangential Direction Approach (TDA), Boundary Fitting Approach (BFA), Tangent Fitting Approach (TFA), and Computerized Tomography Approach (CTA), have been developed to address this problem \citep{wang2022methods,wharton2025modeling}.

To investigate SWCX emissions in geospace, magnetohydrodynamic (MHD) models have been widely used, beginning with the first such effort by \citet{robertson2006x}. The response of SWCX X-ray emission has been studied using MHD simulations \citep{sun2019soft}, compared with XMM-Newton observations \citep{whittaker2016modeling}, and has been investigated alongside other imaging techniques such as energetic neutral atom (ENA) imaging \citep{connor2021soft}. Several studies have also used MHD simulations to derive the magnetopause position from synthetic SWCX observations \citep{collier2018magnetopause, SI8405-Gonzalo, guo2022deriving, kim2023estimating, samsonov2022finding, samsonov2026finding}. These models are computationally efficient because they treat plasma as a fluid, allowing simulations to be performed with relatively high spatial resolution and reasonable computational cost. However, MHD models have limited ability to capture kinetic particle effects and to distinguish between different ion species, despite the ion species dependence of SWCX emission \citep{koutroumpa2023solarwindioncharge}. In addition, their single-fluid formulation does not distinguish between plasma originating from the solar wind and that of magnetospheric origin.  \citet{grandin2024hybrid} used a hybrid-Vlasov approach to investigate soft X-ray emissions at Earth's dayside magnetospheric boundaries, while \citet{guo2023global} used hybrid-PIC simulations to estimate the soft X-ray emission from the terrestrial magnetosheath under different solar wind conditions. Hybrid simulations and test-particle methods provide a better description of particle dynamics, although they are considerably more computationally expensive. Combining MHD simulations with test-particle methods enables simulation of both large-scale plasma dynamics and particle interactions. \citet{xu2024modeling} successfully coupled their LaTeP particle model with MHD simulations to model SWCX emissions and later used these frameworks to investigate X‑ray emissions at the dayside magnetopause under time‑varying solar wind conditions \citep{xu2025dynamic}. Such models therefore promise to combine kinetic physics with the efficiency of global MHD codes to provide a realistic simulation counterpart that can be used to interpret the SMILE SXI dataset.

This paper presents the initial results from Gorgon and the XIM (X-ray Imaging of the Magnetosphere), which combines test-particle simulations with a global MHD model to simulate SWCX emission. We first analyse individual particle trajectories before extending the simulations to particle ensembles to investigate spatial variations in the resulting SWCX emission and evaluate the model's performance. We focus on particle and kinetic effects and compare these results with the pure MHD outputs analysed previously. The simulations capture species-specific spatial variations in SWCX emission, allowing these effects to be investigated. The results are relevant to the SMILE mission and provide a basis for comparison with future observations. 

The remainder of this paper is organized as follows. Section~\ref{sec:methods} describes the Gorgon-XIM model and the simulation methodology. Section~\ref{sec:results} presents the simulation results and compares them with the MHD model. Finally, Section~\ref{sec:conclusions} summarizes the main findings and conclusions.
 
\section{Methods}\label{sec:methods}

\subsection{Global MHD}
 
{Magnetohydrodynamic models have been developed for space plasma applications for two main reasons. First, they provide a global description of regions that cannot be fully characterised using in situ or remote observations alone. Second, they enable theoretical models to be tested and refined by providing a quantitative description of space plasma phenomena \citep{wang2013magnetohydrodynamics}.} 

The Gorgon 3D MHD code was originally developed for high-energy-density laboratory plasma applications \citep{chittenden2004x, ciardi2007evolution} and was later adapted for studies of planetary magnetospheres \citep{mejnertsen2016global,desai2021interplanetary,eggington2022time}. The code maintains a divergence-free magnetic field to machine precision, with separate ion and electron energy equations allow the two species to remain out of thermal equilibrium.
\citet{desai2021drift} coupled Gorgon MHD with test-particle simulations to investigate drift orbit bifurcation dynamics in the outer radiation belt.

In the present study, the MHD equations are solved on a Cartesian grid extending from $-30\,R_{e} \leq X \leq 60\,R_{e}$, $-30\,R_{e} \leq Y \leq 30\,R_{e}$, and $-30\,R_{e} \leq Z \leq 30\,R_{e}$ in Geocentric Solar Magnetospheric (GSM) coordinates, with the Sun located in the negative $X$ direction. The upstream solar wind conditions are specified as $V_{X}=400~\mathrm{km\,s^{-1}}$, $T_{i}=T_{p}=5~\mathrm{eV}$, $B_{z}=-2~\mathrm{nT}$, and proton number density $n_{p}=5~\mathrm{cm^{-3}}$ (see Table~\ref{tab_1}). The inner boundary is located at $3\,R_{e}$ where the plasma density is set to a low value (\(2\times10^{7}~\mathrm{cm^{-3}}\)) to avoid this cold plasma contributing to and artificially affecting the SWCX calculations \citep{walsh2016density}. This simulation configuration produces a subsolar magnetopause near $-10\,R_{e}$ and subsolar bow shock near $-12\,R_{e}$.

\begin{table}[ht]
\centering
\caption{\textbf{Solar wind conditions for simulations}}
\vspace{2mm}
\label{tab_1}
\begin{tabular}{|l|c|c|c|}
\hline
Solar wind type & Density (cm$^{-3}$) & $B_z$ (nT) & Velocity ($V_{x}$) (km\,s$^{-1}$) \\
\hline
Slow & 5 & -2 & 400 \\
\hline
\end{tabular}
\end{table}

Based upon the MHD approximation, Solar wind charge exchange (SWCX) emission is calculated following \citet{cravens1997swcx}, where the local emissivity is given by
\begin{equation}\label{eq_2}
    P = \alpha_{cx} n_{H} n_{sw} V_{rel}
    \quad (\mathrm{eV\,cm^{-3}\,s^{-1}})
\end{equation}

Here, \(\alpha_{cx}\) is an efficiency factor that depends on the solar wind ion composition, charge-exchange cross sections, and branching ratios, typically ranging from \(6\times10^{-16}\) to \(6\times10^{-15}\), and \(n_{H}\) is the neutral hydrogen density of the exosphere. The solar wind proton density, \(n_{sw}\), and relative velocity, \(V_{rel}\), are obtained from the global plasma environment.

The upstream solar wind density, velocity, and interplanetary magnetic field are imposed at the upstream boundary, providing the driving conditions for the simulation. The resulting MHD outputs, particularly the plasma density and velocity, are then used directly to evaluate Equation~(\ref{eq_2}), thereby coupling the global plasma dynamics to the SWCX emissivity.

\subsection{Test Particles}

While fluid models are limited to describing the macroscopic properties of plasmas through the lower-order moments of the particle distribution function, kinetic models provide detailed information on particle dynamics. However, their computational complexity generally limits the spatial and temporal scales that can be modelled using kinetic approaches \citep{swift1996use, winske2003hybrid, von2014vlasiator}. Test-particle methods provide a useful bridge between these two approaches. Using electromagnetic fields obtained from macroscopic models, they can be applied to investigate kinetic effects in complex systems under realistic conditions and have been used to study a broad range of problems in space physics and astrophysics \citep{marchand2010test}.

We employ a forward Lagrangian test-particle approach, in which individual particles (heavy ions) are propagated through the time-dependent electromagnetic fields obtained from the Gorgon MHD simulation. Particle trajectories are advanced using the standard leap-frog \citet{Boris70} scheme, which is widely used to integrate the motion of charged particles in electromagnetic fields. This method preserves phase-space volume and provides stable and accurate integration of particle gyromotion. 

The test particle simulations and MHD simulations are fully integrated and performed in line together \citep{desai2021interplanetary}, within a smaller particle simulation domain that extends from $-20\,R_e \leq X \leq 0\,R_e$, $-10\,R_e \leq Y \leq 10\,R_e$, and $-10\,R_e \leq Z \leq 10\,R_e$. All simulations are performed under the quiet solar wind conditions summarized in Table~\ref{tab_1}. The remaining upstream parameters are specified as $V_{y}=V_{z}=0$ and $B_{x}=B_{y}=0$. A total of $2\times10^{7}$ particles are initialized, per ion species, within the upstream solar wind, to ensure statistically robust results. Each particle is initialized with the bulk solar wind velocity together with an additional random thermal component to represent a thermal distribution.

\begin{figure}[h!]
     \begin{center}
             \includegraphics[width=0.85\textwidth]{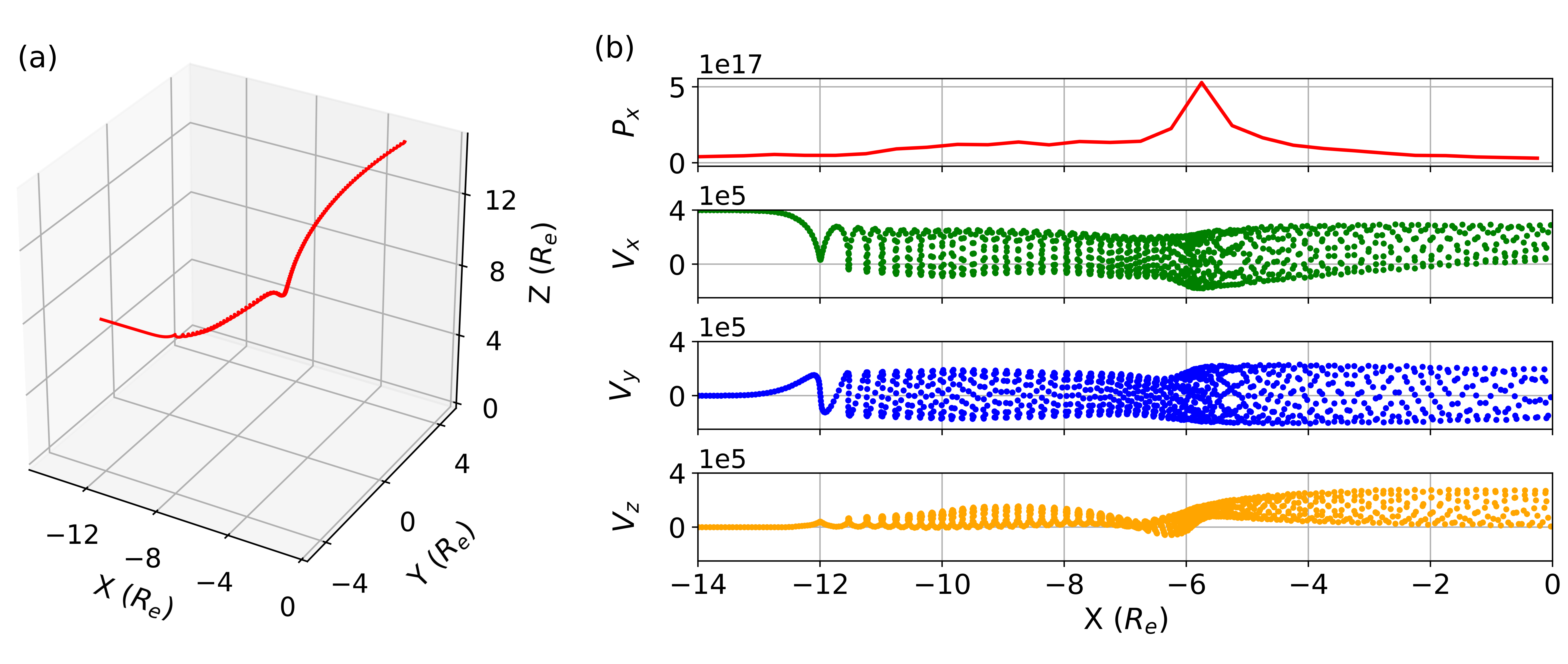}
     \end{center}
     \caption{(a) Example of a single ion trajectory in the simulation. (b) The change in variables such as $P_x$ (X-ray emission), $V_x$, $V_y$, $V_z$ (X, Y and Z components of velocity) across the particle trajectory.}
      \label{fig_1}
 \end{figure}

The MHD simulation is evolved for 120 minutes before particle injection to allow the background electromagnetic fields to reach a quasi-steady state. The model is first validated using single-particle trajectories, allowing a detailed examination of particle motion through key regions such as the bow shock and magnetosheath. Figure~\ref{fig_1} shows the trajectory of a representative test particle. In the upstream solar wind, the electric and magnetic forces largely balance in the solar wind frame, resulting in an approximately straight-line trajectory. Upon encountering the bow shock at approximately $-12\,R_{e}$, where the magnetic field strength increases to roughly four times its upstream value, the particle begins to undergo pickup-ion-like acceleration, with its velocity oscillating between the solar wind speed and approximately twice the magnetosheath flow speed differential. The particle follows this characteristic trajectory as it enters the cusp near $-6\,R_{e}$ before being transported downtail. This repeated acceleration is characteristic of the magnetospheath and contributes to the effective temperature herein.

Convergence tests are performed to ensure that the selected numerical parameters and timestep provide accurate results while maintaining computational efficiency. Following validation, the full particle population is evolved within the MHD fields.

\subsection{Gorgon-XIM}

The inclusion of test particles enables a modified treatment of solar wind charge exchange (SWCX) emission, incorporating particle-specific properties and interaction losses. The emissivity expression (Equation~\ref{eq_2}) is therefore extended to:

\begin{equation}\label{eq_3}
    P = \alpha_{cx} n_H n_{X^{q+}} V_{rel} \quad (eV \, cm^{-3} \, s^{-1}),
\end{equation}

where $n_H$ is the exospheric density and $n_{X^{q+}}$ represents the time-varying density of the test particle species, accounting for losses due to charge exchange interactions.

In this formulation, the efficiency factor \(\alpha_{cx}\) is not treated as a constant but is evaluated as a function of ion species and emission processes. For a given ion \(X^{q+}\), it is expressed as:

\begin{equation}\label{eq_4}
    \alpha_{cx} = \sum_{X^{q+}} \sum_{j} \sigma_{X^{q+}, M} \,[X^{q+}/\mathrm{O}] \,[\mathrm{O}/\mathrm{H}] \, E_{j} \, Y_{X^{(q-1)+}, j},
\end{equation}

where \(\sigma_{X^{q+}, M}\) denotes the charge exchange cross-section between the solar wind ion \(X^{q+}\) and neutral species \(M\). The ratios \([X^{q+}/\mathrm{O}]\) and \([\mathrm{O}/\mathrm{H}]\) represent solar wind ion abundances, \(E_{j}\) is the photon energy associated with transition \(j\), and \(Y_{X^{(q-1)+}, j}\) is the corresponding branching ratio.

Charge exchange cross-sections used in this study are compiled from a range of experimental and empirical sources, as no single dataset provides complete coverage for all relevant ion species and energies. The cross sections are velocity dependant the adopted values are summarized in Table~\ref{tab_2}. The heavy ion composition of the solar wind is a key parameter in determining SWCX emission, as it reflects the source region, coronal temperature, and solar wind type \citep[e.g.,][]{zhao2009swtype, nittin2024swtype}. Measurements from the Advanced Composition Explorer (ACE), in particular the Solar Wind Ion Composition Spectrometer (SWICS) instrument \citep{gloeckler1998ACE}, have been widely used to characterize solar wind heavy ion abundances and charge states. These datasets have been extensively applied in SWCX studies \citep[e.g.,][]{koutroumpa2023solarwindioncharge}. Following the loss of heavy-ion measurements from SWICS in 2011, newer missions such as the Interstellar Mapping and Acceleration Probe (IMAP) \citep{livi2026IMAP} may provide new upstream measurements, while planned missions could measure heavy-ion abundances directly adjacent to the magnetopause \citep{Carter2026Elfen}. For the present study, representative abundance values consistent with quiet solar wind conditions are adopted.

\begin{table}[htbp]
\centering
\caption{\textbf{Charge exchange parameters for selected ions}}
\vspace{2mm}
\begin{tabular}{|c|c|c|c|c|c|}
\hline
Ion ($X^{q+}$)
& Mass per 
& $\sigma_{\alpha}$
& $X^{q+}/O$ 
& $E_{j}$ 
& $Y_{j}$ \\
& charge
&($10^{-15}\,\mathrm{cm}^{2}$)
&
&($\mathrm{eV}$)
& \\
\hline
$C^{5+}$   & 2.40  & 2.00 & 0.210 & 299 & 0.838\\
$C^{6+}$   & 2.00  & 4.16 & 0.318 & 367 & 0.438\\
$Mg^{11+}$ & 2.18  & 7.50 & 0.035 & 1330 & 0.262\\
$N^{6+}$   & 2.33  & 3.71 & 0.058 & 420 & 0.587\\
$N^{7+}$   & 2.00  & 5.67 & 0.006 & 500 & 0.437\\
$Ne^{9+}$  & 2.22  & 7.20 & 0.030 & 922 & 0.141\\
$O^{7+}$   & 2.28  & 3.40 & 0.200 & 561 & 0.473\\
$O^{8+}$   & 2.00  & 5.65 & 0.070 & 653 & 0.443\\
\hline
\end{tabular}
\vspace{2mm}

\footnotesize
Source: \citet{bodewits2007spectral, koutroumpa2006charge, cumbee2021kronos, beijers1994state, bliman1992single, bonnet1985electron, dijkkamp1985selective, dijkkamp1985subshell, fritsch1996one, greenwood2001experimental, harel1992double, harel1998cross, ishii2004electron, iwai1982cross, lee2004charge, liu2005charge, phaneuf1987heavy, richter1993application, shimakura1992molecular, suraud1991state, wu1994evidence, garcia1965energy, vainshtein1985energy, drake1988theoretical, savukov2003multipole, dere1997chianti, landi2006chianti} 
\label{tab_2}
\end{table}

The oxygen-to-hydrogen ratio is taken as \([\mathrm{O}/\mathrm{H}] = 4.76 \times 10^{-4}\), representative of slow solar wind conditions \citep{Lepri2013OHratio}. Variations in this ratio, associated with different solar wind regimes, can significantly influence \(\alpha_{cx}\) and the resulting emission \citep[e.g.,][]{Liang2023alphaswcx, nittin2024swtype}.

Branching ratios and transition data are obtained from the KRONOS database \citep{cumbee2021kronos}, which provides velocity-dependent charge exchange cross-sections, energy levels, and line emission data based on multichannel Landau–Zener (MCLZ) calculations. The ion species included in this study are $C^{5+}$, $C^{6+}$, $Mg^{11+}$, $N^{6+}$, $N^{7+}$, $Ne^{9+}$, $O^{7+}$, and $O^{8+}$, selected based on their contributions to observed SWCX emission in X-ray observations \citep[e.g.,][]{jenny2009XMM}.

The neutral hydrogen density is approximated as

\begin{equation}\label{eq_5}
    n_H = 25 \left(\frac{10 \,R_e}{r}\right)^3
\end{equation}

where \(R_e\) is Earth’s radius and \(r\) is the radial distance \citep{cravens2001swcx}. Although simplified, this expression captures the radial dependence of exospheric hydrogen density, which strongly influences SWCX emission intensity. Enhanced emission is expected in regions such as the cusps, where neutral densities are elevated. The exospheric density is sensitive to solar wind–magnetosphere–atmosphere coupling processes, including photoionization, charge exchange, and atmospheric upwelling, although accurate modelling remains limited by the availability of in-situ measurements \citep[e.g.,][]{Connor2021nHdensity}.

To account for ion depletion during charge exchange, each test particle is assigned a weight that decreases over time \citep{desai2021photodetachment}. This is implemented as

\begin{equation}\label{eq_6}
    n_{X^{q+}}(t) = n_{X^{q+}}(t - \Delta t)\, e^{-k \Delta t}.
\end{equation}

This formulation captures the progressive reduction in ion population as particles undergo successive interactions. 

An additional advantage of using test particles with time-dependent weights is that the framework can be extended to model charge exchange cascades. For example, an $O^{8+}$ ion can undergo charge exchange with a neutral atom, producing an excited $O^{7+}$ ion and an X-ray photon, as described by Equation~\ref{eq_1}. This process can be represented by decreasing the weight of the $O^{8+}$ test particle and increasing the weight of an $O^{7+}$ test particle at the same location. The resulting $O^{7+}$ ions can subsequently undergo further charge-exchange interactions, producing lower charge states such as $O^{6+}$. The time-dependent particle weights therefore provide a framework for tracking multiple charge-exchange interactions and their associated cascades throughout the simulation. However, this cascading effect is not included in the present study.

All parameters in the model can vary with solar wind and magnetospheric conditions. The SWCX emissivity is calculated using Equation~(\ref{eq_3}) and used to generate the resulting emission distributions.

\section{Results} \label{sec:results}

\subsection{SWCX Maps}

Following the completion of the simulations, synthetic X-ray emission maps are generated for both the MHD-only case and the combined MHD–test particle model. To enable a direct comparison, a ``particle mask'' is applied to the MHD results, such that regions with zero particle contribution in the coupled XIM simulation are also excluded in the MHD-only maps.

\begin{figure}[h!]
     \begin{center}
             \includegraphics[width=0.85\textwidth]{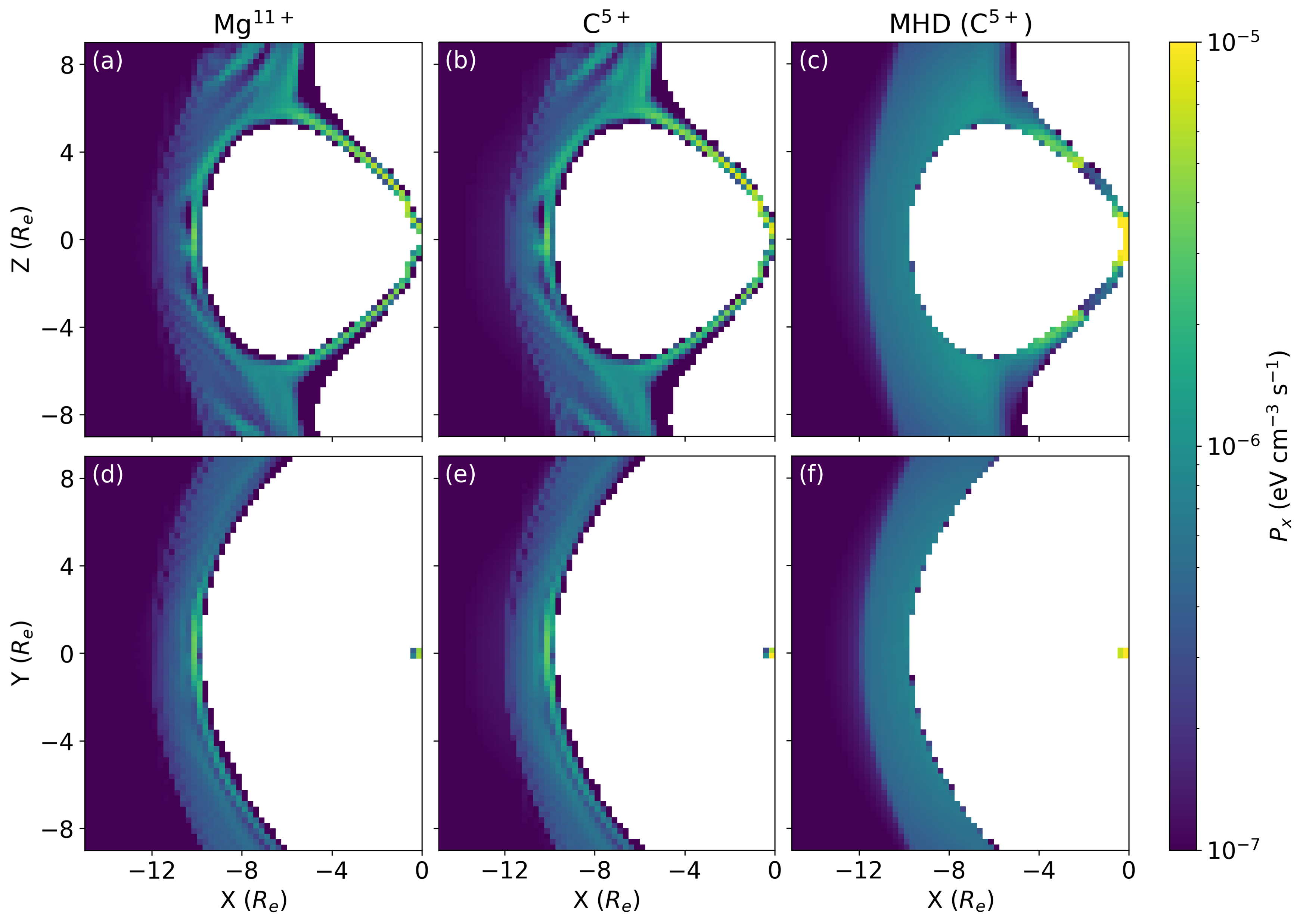}
     \end{center}
     \caption{SWCX emission maps. Panels (a)--(c) show emissions in the X--Z plane ($Y=0$), while panels (d)--(f) show emissions in the X--Y plane ($Z=0$).}
      \label{fig_2}
\end{figure}

Figure~\ref{fig_2} presents two-dimensional slices of the emission for the ion species $Mg^{11+}$ and $C^{5+}$, along with the corresponding MHD-only emission for $C^{5+}$. The top row shows slices in the X--Z plane ($Y = 0$), while the bottom row shows slices in the X--Y plane ($Z = 0$), both passing through the center of the Earth. Key magnetospheric structures, including the bow shock and magnetopause, are clearly resolved in all cases. Enhanced emission is observed in the cusp regions, particularly in the X--Z slices. While the overall spatial distribution of emission is similar between different ion species, noticeable differences arise when comparing the MHD-only and coupled MHD–test particle results.

The MHD-only emission appears comparatively smooth, whereas the inclusion of test particles introduces additional structure, particularly within the magnetosheath. These features are associated with localized variations in particle dynamics that are not captured by the fluid approximation alone. The enhanced emission within the magnetosheath may be related to particle heating associated with increasing magnetic field strength. Despite these differences in spatial structure, the overall emission intensity remains of the same order of magnitude in both approaches.
An increase in emission intensity is also observed closer to the Earth in all simulations, particularly near the cusp regions, consistent with higher neutral hydrogen densities and enhanced charge exchange rates in these areas.

\subsection{LOS Images}

\begin{figure}[ht]
     \begin{center}
             \includegraphics[width=0.85\textwidth]{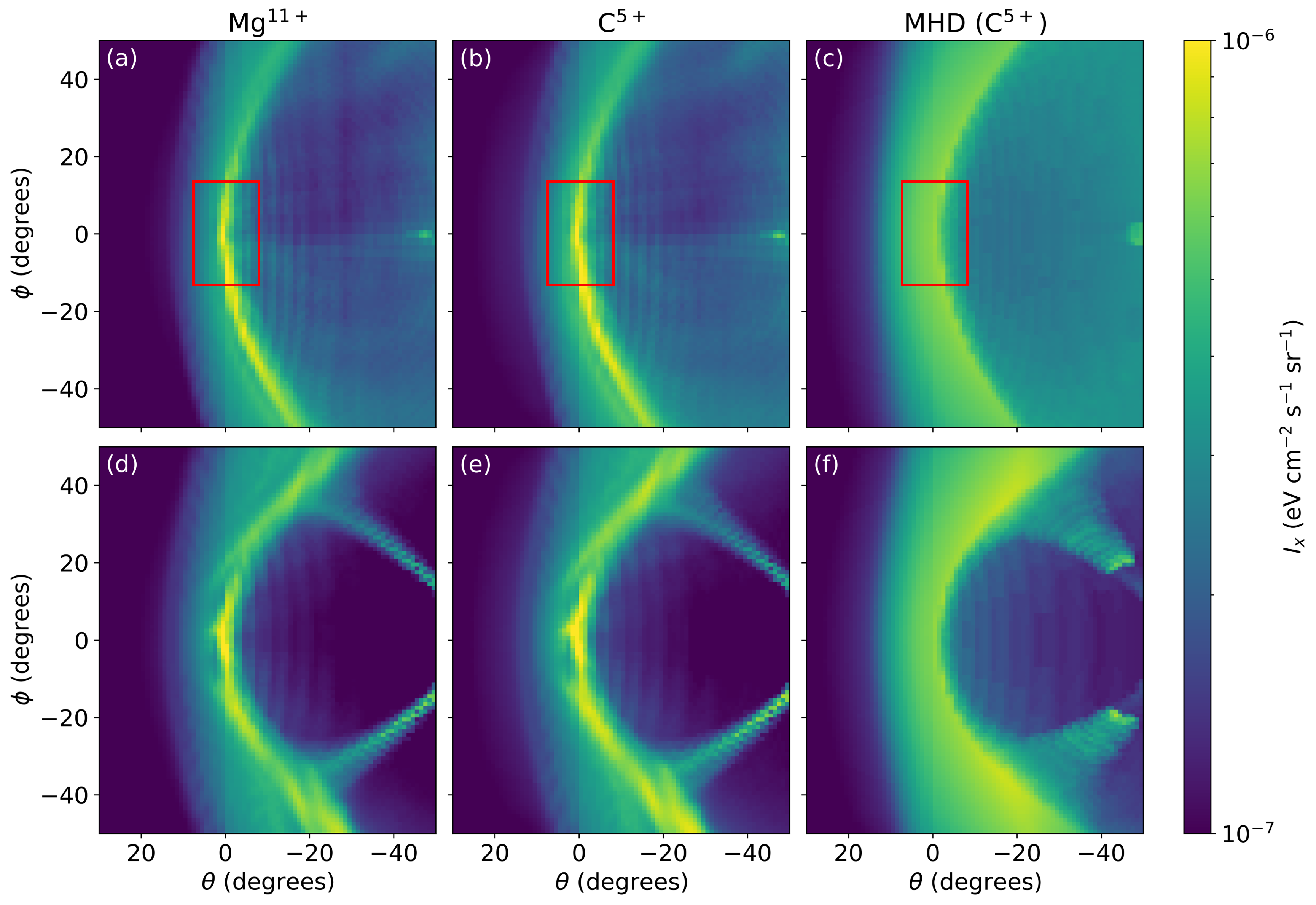}
     \end{center}
     \caption{Line-of-sight (LOS)-integrated SWCX emission maps. Panels (a)--(c) show the results for $Mg^{11+}$, $C^{5+}$, and the MHD-only emission for $C^{5+}$, respectively, for satellite viewpoint 1. The red box in these panels represents the SMILE field of view (FOV) for reference. Panels (d)--(f) show the corresponding results for satellite viewpoint 2.}

      \label{fig_3}
 \end{figure}

To facilitate comparison with satellite observations, line-of-sight (LOS)-integrated X-ray images are generated from the simulations. Figure~\ref{fig_3} presents synthetic LOS images for the ion species $Mg^{11+}$ and $C^{5+}$, together with the corresponding MHD-only emission for $C^{5+}$. (For a detailed description of the LOS calculations, see the Appendix) In the top three panels, the synthetic satellite is located at $(-10\,R_e, 0\,R_e, 10\,R_e)$ (viewpoint 1), while in the bottom three panels it is located at $(-10\,R_e, 10\,R_e, 0\,R_e)$ (viewpoint 2).

The images clearly show large-scale magnetospheric structures, including enhanced emission from the cusp regions, as well as contributions from the magnetopause and magnetosheath. Due to the viewing geometry in viewpoint 1, where the system is observed from above, the two cusp regions are not easily distinguishable and appear as a single emission feature. In contrast, the cusp structures are more clearly resolved in viewpoint 2.
The overall spatial distributions of $Mg^{11+}$ and $C^{5+}$ are similar. However, as seen in the planar slices, the inclusion of test particles introduces additional structure in the emission. These features are particularly evident when compared with the smoother MHD-only results, highlighting the influence of particle dynamics on the SWCX process.

The emissions in the cusp regions are significantly different between the MHD only and MHD with test-particle simulations. The MHD-only simulation produces comparatively stronger emission in the cusp regions, consistent with the higher plasma densities predicted by the fluid model. In contrast, the test-particle emission is much narrower and more strongly peaked towards the centre of the cusp, with the emission decreasing towards its edges. This structure reflects the allowed trajectories of the test particles and the regions in which the particles are able to contribute to the SWCX emission. Although a particle mask is applied to the MHD emission, when a particle is present, the emission is still calculated using the MHD plasma density and temperature. Consequently, the MHD-only emission produces a broader emission structure across the full width of the cusp. This difference also affects the radial structure of the cusp emission. For example, in Figure~\ref{fig_2}, very high emission is observed close to the Earth. This may include contamination from ionospheric plasma \citep{walsh2016density}, while the MHD plasma emission peaks at the inner boundary of $3\,R_E$. The test particles, on the other hand, can reach realistic mirror points inside the inner boundary, allowing the resulting emission to exhibit a different radial structure.

Despite these differences, the synthetic images show good qualitative agreement between the modelling approaches and capture the key features expected in soft X-ray observations.
These results demonstrate that the modelling framework is capable of producing realistic synthetic observations from different viewpoints and can be adapted for direct comparison with satellite measurements.

\subsection{Effect of Losses on Spatial Variations}

As summarized in Table~\ref{tab_2}, each heavy ion species exhibits a different charge exchange cross-section, mass per charge ratio, and  with exospheric hydrogen, leading to species-dependent loss rates. These differences directly influence the spatial distribution of SWCX emission, particularly within the magnetosheath where interactions are most significant.

\begin{figure}[ht]
     \begin{center}
             \includegraphics[width=0.85\textwidth]{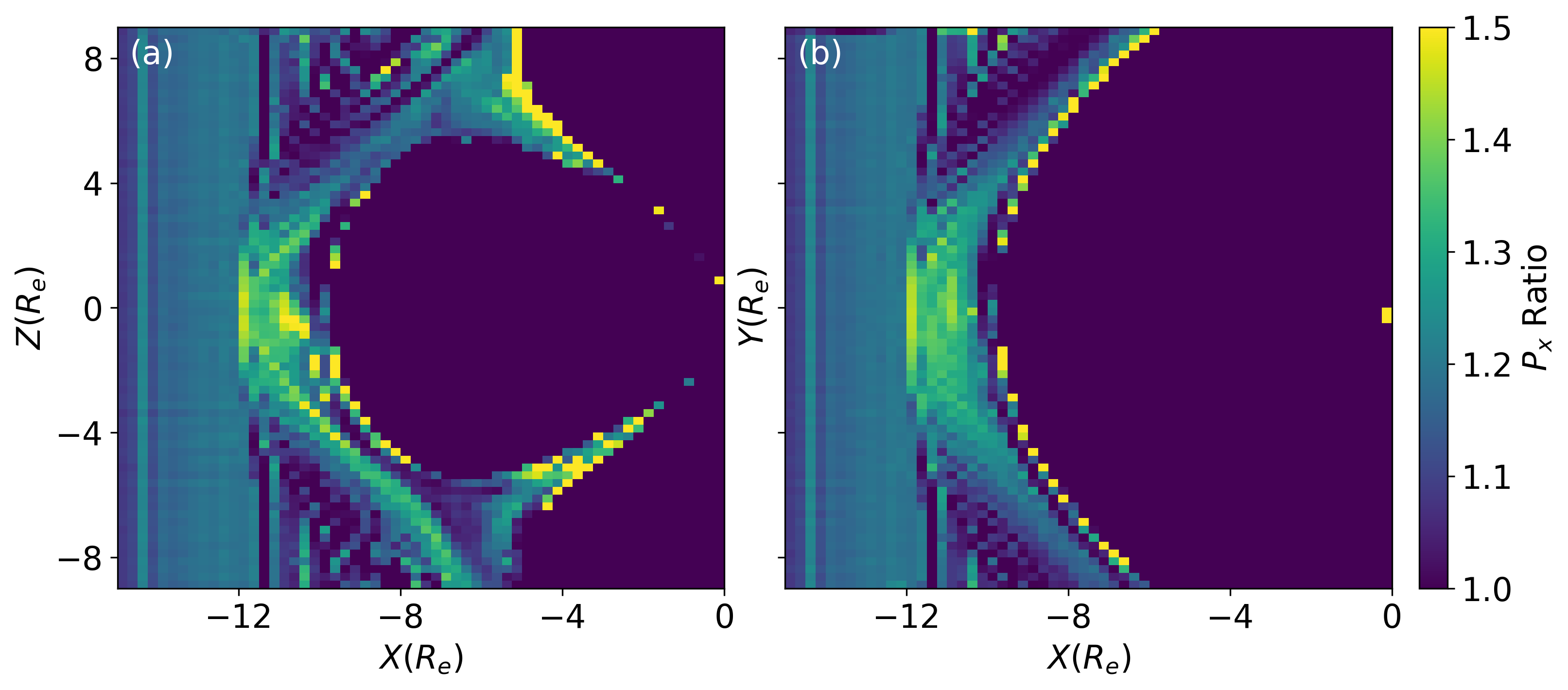}
     \end{center}
     \caption{The ratio of emission of $C^{5+}$ to $Mg^{11+}$ has been calculated and plotted. (a) Shows the 2D sliced image of the same in X-Z plane, where Y=0 and (b) Shows the 2D sliced image of the same in X-Y plane, where Z=0.}
      \label{fig_4}
 \end{figure}

To investigate this effect, we compare the emission from two representative ion species, $Mg^{11+}$ and $C^{5+}$. The ion $Mg^{11+}$ has a relatively large charge exchange cross-section ($\sigma \approx 7.50 \times 10^{-15}$ cm$^{2}$), whereas $C^{5+}$ has a significantly smaller value ($\sigma \approx 2.00 \times 10^{-15}$ cm$^{2}$). The two species are also well separated in energy at the extremes of the energy range of interest. This contrast allows us to isolate the impact of interaction losses on the resulting X-ray emission.
Figures~\ref{fig_4}(a) and (b) show the spatial distribution of the emission ratio between $C^{5+}$ and $Mg^{11+}$. In the upstream solar wind region, the emission ratio remains close to $\sim 1.25$, indicating comparable emission levels as theoretically predicted by Equation \ref{eq_3} for the nominal solar wind values.

\begin{figure}[ht]
     \begin{center}
             \includegraphics[width=0.85\textwidth]{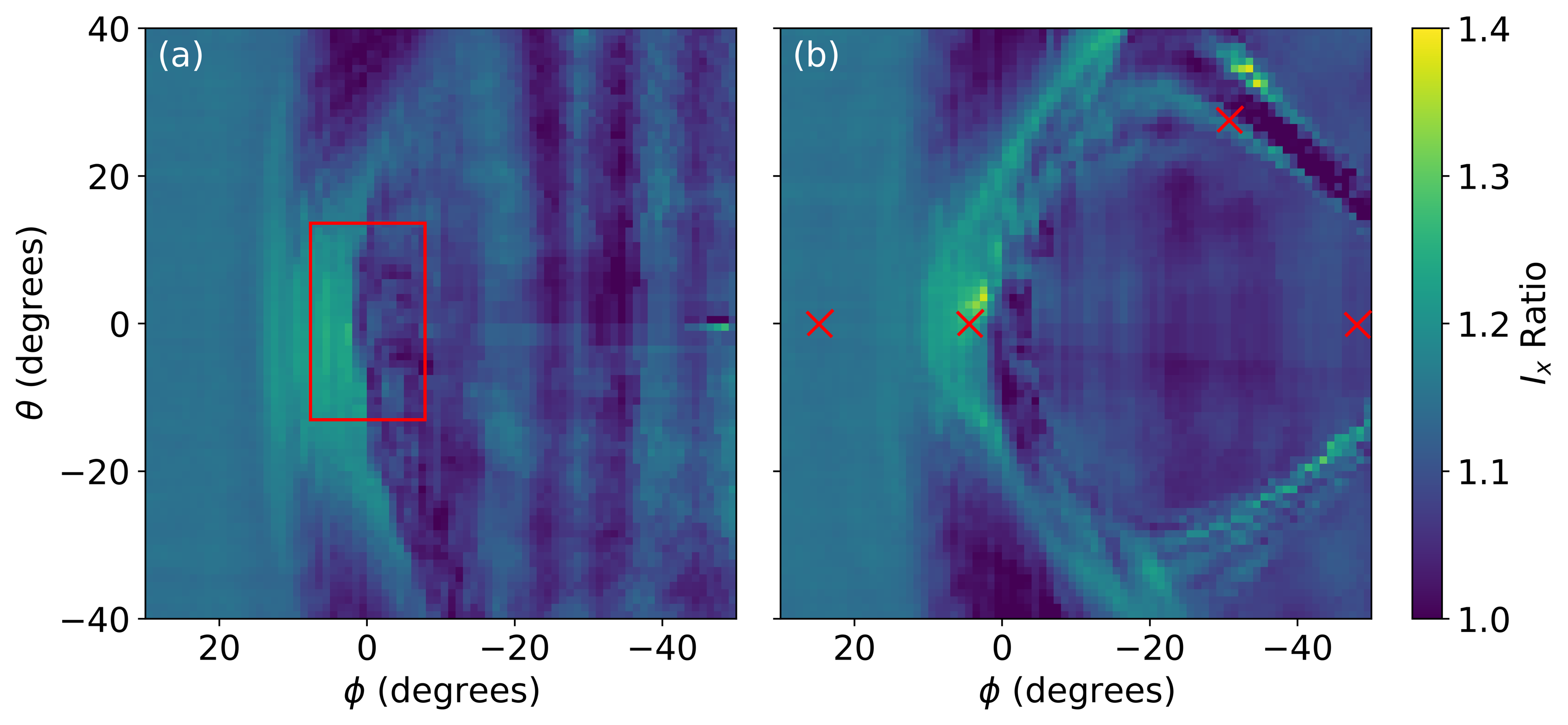}
     \end{center}
     \caption{Line-of-sight (LOS) integrated SWCX emissions ratio of $C^{5+}$ to $Mg^{11+}$ has been calculated and plotted. (a) Shows the integrated emission from viewpoint 1. (b) Shows the integrated emission from viewpoint 2. The red box represents the SMILE FOV.}
      \label{fig_5}
 \end{figure}

However, as the plasma passes through the bow shock and into the magnetosheath, the ratio increases. Figure \ref{fig_4} shows a single slice through the simulation and shows large localised spatial variations, potentially associated with the different trajectories of the different ions. The large-scale effects are therefore best viewed in Figure \ref{fig_5} showing the line-of-sight (LOS) integrated ratio map. Here the ratio increases from 1.25 to around 1.3 near the magnetopause. We believe this enhancement where the interaction with the hydrogen geocorona is strongest, reflects the more rapid depletion of $Mg^{11+}$ due to its larger charge exchange cross-section, resulting in a relative increase in the contribution from $C^{5+}$. The effect, while minor, does highlight the importance of species-dependent loss processes in shaping the spatial structure of SWCX emission. Both Figures~\ref{fig_4} and~\ref{fig_5} also show interesting structure in the emission ratio near the edges of the cusp regions, which we associate with differences in their mass per charge ratios and, consequently, the gyroradii and trajectories of the different ion species. Differences are also apparent between the two cusps, and the factors affecting ion entry into the magnetically complicated cusp regions will be investigated within a future study.

\begin{figure}[ht]
     \begin{center}
             \includegraphics[width=0.85\textwidth]{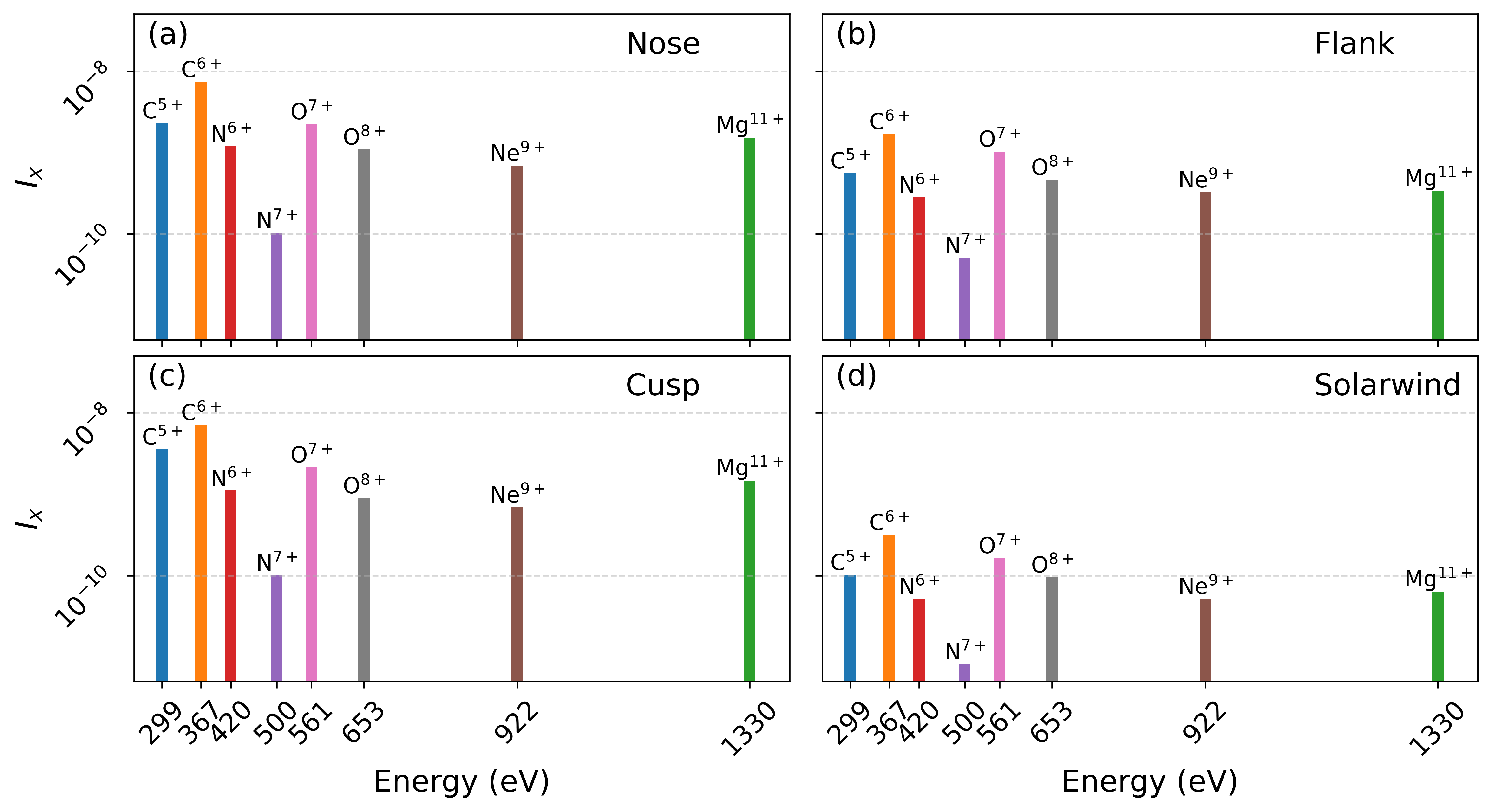}
     \end{center}
     \caption{Panels (a)--(d) show single line-of-sight (LOS) emissions from SMILE located at [-10, 0, -10] looking at the magnetopause nose, flank, cusp, and solar wind, respectively, as viewed from viewpoint 1. These isolated LOS measurements are used to generate charge-exchange emission spectra, enabling comparisons of emissions from different regions and illustrating how they vary under different solar wind types and conditions.}
      \label{fig_6}
 \end{figure}

The influence of species-dependent losses and ion time-history in the spatial distribution of the emission, will therefore influence all ions with the X-ray spectrum.
Figure~\ref{fig_6} presents a preliminary example of the simulated emissivity spectra for different ion species at several locations within the magnetosphere. The selected regions are indicated by the red crosses shown on Figure~\ref{fig_5}(b). Only the major line transitions are shown in the figure, although minor transitions are also present but are not included in the current study. Each species has different charge-exchange cross sections and solar wind abundances, resulting in a species-dependent efficiency factor, $\alpha_{cx}$. In addition, the different charge-to-mass ratios of the ions affect their particle transport and trajectories within the MHD fields. These species-dependent properties therefore influence the charge-exchange emissions and particle transport, even though the external parameters are initially set to be the same. Differences in the relative contributions of individual ion species between locations demonstrate how charge-exchange losses and particle transport modify the observable SWCX emission and may be used as a diagnostic of transport processes. In a physical X-ray instrument, the distinction between these lines will depend on the energy redistribution of the instrument, so sufficiently fine spectral resolution is important for resolving these features.

\section{Conclusions}\label{sec:conclusions}

Within this study, Gorgon-XIM combined the Gorgon MHD model with test-particle simulations to model SWCX emission from Earth's magnetosphere, providing an enhanced framework for predicting the spatial distribution and X-ray signatures of the system and enabling detailed comparisons with global X-ray observations. The coupled model successfully reproduces key magnetospheric regions, including the bow shock, magnetopause, magnetosheath, and cusps. Direct comparison with the corresponding MHD-only simulations shows that Gorgon-XIM captures additional spatial structure that is absent from the fluid model alone. In particular, enhanced emission is observed within the magnetosheath, particularly near the magnetopause, which may reflect particle heating as ions propagate through regions of increasing magnetic field strength. This behaviour appears consistent with betatron heating, where the local magnetic field properties play a role shaping the X-ray emissions through enhancing the relative velocity between the solar wind ions and exospheric neutrals which are blind to the magnetic field. These differences are also evident in the line-of-sight (LOS) images, where enhanced emission is observed from the cusps and magnetopause. Such synthetic LOS images provide a useful framework for comparison with future X-ray observations.

The time history of ions, together with their relative abundances in the solar wind and their charge-exchange cross sections, produces species-dependent variations in SWCX emission throughout the magnetosphere. Synthetic LOS observations were used to generate X-ray spectra for different ion species at different viewing locations. These spectral variations have the potential to provide diagnostic information on plasma dynamics and solar wind conditions \citep{nittin2024swtype}. To investigate these effects, LOS emission ratio maps were produced for the ion species \(C^{5+}\) and \(Mg^{11+}\). The results indicate a more rapid depletion of X-ray emission from \(Mg^{11+}\), consistent with its larger charge-exchange cross section. Such heavy-ion ratio studies may provide an additional means of validating the model against observations, such as those reported by \citet{carter2011identifying}, and may also assist in identifying magnetospheric boundaries, including the bow shock \citep{xu2024modeling}. The results also demonstrate the strong influence of the efficiency factor, \(\alpha_{cx}\), on the resulting SWCX emission. Future work will apply Gorgon-XIM to observed solar wind events and compare the simulated emissions with measurements from missions such as SMILE.

\section*{Acknowledgements} \label{sec:acknowledgements}
AR acknowledges support from the Chancellor’s International Scholarship (2023), funded by the University of Warwick. RTD acknowledges an STFC Ernest Rutherford Fellowship ST/W004801/1 and award ST/Y005635/1. HLN acknowledges support from the EPSRC. YT supported by EPSRC and UK MET office. DB is EPSRC and Qinetiq. SN is supported by the University of Leicester’s Future-100 Scholarship scheme and by Royal Society Grant DHF/R1/211068. JAC gratefully acknowledges funding from a Royal Society Dorothy Hodgkin Fellowship DHF/R1/211068.

\section*{Appendix}
\subsection*{Synthetic Imaging and Line-of-Sight Integration}

To enable comparison with observations from the SMILE Soft X-ray Imager (SXI), synthetic X-ray images are constructed using line-of-sight (LOS) integration of the modeled emissivity. This requires mapping the simulation outputs onto an instrument viewing geometry and performing coordinate transformations between the simulation and image frames.

We consider an idealized X-ray imager located at position
\begin{equation}
\mathbf{r}_{\mathrm{sat}} = (10.0,\, 0.0,\, 10.0)\, R_e,
\end{equation}
pointing toward the target location
\begin{equation}
\mathbf{r}_{\mathrm{target}} = (10.0,\, 0.0,\, 0.0)\, R_e.
\end{equation}

The unit vector along the boresight (viewing direction) is defined as
\begin{equation}
\hat{\mathbf{Y}} = \frac{\mathbf{r}_{\mathrm{target}} - \mathbf{r}_{\mathrm{sat}}}
{\left\lVert \mathbf{r}_{\mathrm{target}} - \mathbf{r}_{\mathrm{sat}} \right\rVert}.
\end{equation}

To construct a right-handed coordinate system for the imager, a reference vector
\begin{equation}
\mathbf{x}_{\mathrm{ref}} = (1,\, 0,\, 0)
\end{equation}
is introduced, from which the horizontal axis is obtained as
\begin{equation}
\hat{\mathbf{X}} = \frac{\mathbf{x}_{\mathrm{ref}} \times \hat{\mathbf{Y}}}
{\left\lVert \mathbf{x}_{\mathrm{ref}} \times \hat{\mathbf{Y}} \right\rVert}.
\end{equation}

The vertical axis is then defined using the right-hand rule:
\begin{equation}
\hat{\mathbf{Z}} = \hat{\mathbf{X}} \times \hat{\mathbf{Y}}.
\end{equation}

The orthonormal basis $(\hat{\mathbf{X}}, \hat{\mathbf{Y}}, \hat{\mathbf{Z}})$ defines the instrument coordinate system and corresponding image plane.

Each pixel in the image is associated with angular coordinates $(\theta, \phi)$ representing deviations from the boresight direction, where $\theta$ denotes the horizontal (azimuthal) angle and $\phi$ the vertical (elevation) angle. For an image of size $N_x \times N_y$ with total field of view $\mathrm{FOV}_x \times \mathrm{FOV}_y$, the angular sampling is given by
\begin{equation}
\theta_i = \left(i - \frac{N_x}{2}\right)\Delta\theta, \quad
\phi_j = \left(j - \frac{N_y}{2}\right)\Delta\phi,
\end{equation}
with
\begin{equation}
\Delta\theta = \frac{\mathrm{FOV}_x}{N_x}, \quad
\Delta\phi = \frac{\mathrm{FOV}_y}{N_y}.
\end{equation}

For each pixel $(i,j)$, the corresponding line-of-sight unit vector is constructed as
\begin{equation}
\mathbf{v}_{\mathrm{LOS}} = 
\frac{\hat{\mathbf{Y}} + \theta_i \hat{\mathbf{X}} + \phi_j \hat{\mathbf{Z}}}
{\left\lVert \hat{\mathbf{Y}} + \theta_i \hat{\mathbf{X}} + \phi_j \hat{\mathbf{Z}} \right\rVert}.
\end{equation}

The LOS path is parameterized as
\begin{equation}
\mathbf{r}(s) = \mathbf{r}_{\mathrm{sat}} + s\, \mathbf{v}_{\mathrm{LOS}},
\end{equation}
where $s$ denotes the distance along the ray.

The simulation domain is discretized with grid spacing $\Delta = 0.25\, R_e$. Each LOS is sampled at intervals
\begin{equation}
s_k = k \Delta, \quad k = 0,1,2,\dots
\end{equation}
until the ray exits the simulation domain.

The X-ray intensity for each pixel is computed via line integration of the emissivity $P_X$:
\begin{equation}
I_X = \frac{1}{4\pi} \int P_X(\mathbf{r}) \, ds \;\approx\;
\frac{1}{4\pi} \sum_k P_X\big(\mathbf{r}(s_k)\big)\, \Delta.
\end{equation}

This procedure is repeated for all pixels to construct the synthetic X-ray image.

For the results presented in this study, a field of view of $100^\circ \times 100^\circ$ is adopted to provide a comprehensive visualization of the global emission structure. The SMILE SXI instrument has a smaller field of view of $15.5^\circ \times 26.5^\circ$ \citep{sembay2024soft, sembay2025soft}; however, the model is fully adaptable to this observational configuration for direct comparison with mission data.

\bibliography{bibliography}

\end{document}